\documentclass[12pt]{article}

\usepackage{amsmath,amsfonts,amsbsy,amssymb,fullpage,makeidx,array,graphicx,epsfig,enumerate}

\makeindex
\newcommand{\phiav}{\langle\phi\rangle}

\newcounter{defncntr}

\newcounter{thmcntr}

\newcounter{remcntr}

\renewcommand{\theremcntr}{\arabic{section}.\arabic{remcntr}}

\newcounter{lemcntr}

\newcounter{egcntr}

\begin{document}

\thispagestyle{empty}

\title{Of Inflation and the Inflaton}

\author{ R. Brout\thanks{Essay written for the Gravity Research
Foundation 2010 Awards for Essays on Gravitation}
\\
\small Department of Applied Mathematics, University of Waterloo\\
\small Waterloo, Ontario N2T 3G1, Canada\\[8mm]
\small Service de Physique Th\'eorique, Universit\'e Libre de Bruxelles \\
\small The International Solvay Institutes\\
\small B1050 Bruxelles, Belgium\\ \small robert.brout@ulb.ac.be}

\date{29 March 2010}

\maketitle


\begin{abstract}
  \noindent  Due to intra-field gravitational interactions, field
  configurations have a strong negative component to their energy
  density at the planckian and transplanckian scales, conceivably resulting in
  a sequestration of the transplanckian field degrees of freedom.
  Quantum fluctuations then allow these to tunnel into
  cisplanckian configurations to seed inflation and conventional observed
  physics: propagating modes of QFT in a geometry which responds to
  the existence of these new modes through the energy constraint of
  general relativity, $H^2 = \rho/3$. That this tunnelling results in
  geometries and field configurations that are homogeneous
  allows for an estimate of the mass of the
  inflaton, $m=O(10^{-6})$, and the amplitude of the inflaton
  condensate, $\phiav=O(10)$, both consistent with
  phenomenology.

\end{abstract}

\newpage
\section{Introduction}

This paper is an enquiry into the conceptual foundations of
cosmogenesis as it is suggested within the inflationary scenario.  In
particular, the highly successful implementation of inflation through
the dynamics of the inflaton field ($\equiv \phi$) will be discussed.

It is not without interest to brook the subject as it was conceived in the
creation scenario of early civilizations.  The first lines of Genesis 1.1.,
thoughts of the ancient Hebrews, believed to date from 700 B.C. or thereabouts,
read:

\begin{verse}
  ``And the earth was without form and void and darkness was
  on the face of the deep.\\
  And God said, Let there be light and there was light...''
\end{verse}

\noindent Similar notions are founds in the clay tablets of the more urbanized
Sumerians and Babylonians from a millennium beforehand.

That the universe is a manifestation of form created from formless
void is a natural thought.  One has to begin somewhere.

However the second part of the proposition has now been seriously
called into question. ``Let there be light'' implies suddenness.  It
corresponds to the big bang as it was conceived before inflation, the
existence of an initial state of homogeneity on a space-like surface
from which the cosmos was born, the inception, unexplained, of the era
of adiabatic expansion.  It took some years for relativity to
penetrate our collective scientific consciousness to abandon this view.
 The universe, concomitant with its homogeneity in the mean,
must result from a cause, the sudden onset of homogeneity rejected in
favor of its gradual establishment from prior circumstances.  Just as
action at a distance must be abandoned, the world, in all its aspects,
is to be concerned as a succession of events from some initial event
onwards.  Short of this, our only other resource would indeed be
divine intervention.

We shall follow historical precedent and adopt biblical chaos as the
precondition of cosmology, in a sense to be more precisely described.
General relativity (GR) suggests that the initial seed
 of creation was planckian in dimensions.
From that seed, a background geometry, including time itself, must emerge.
 Quantum field theory (QFT) incites us to seek, within the seed, the birth
 of field configurations which serve as the basis of physics as we practice it,
the field theoretical modes. Inflation ensues wherein field configurations,
modes of quantum fields, propagate in geometry; and geometry, through
the constraints of GR, feeds on the energy of the modes. The requirement
of causality has led us that far. Inflation is the child of relativity and
through the two modern instruments of physics,
GR and QFT, we are called upon to bring it to maturity.

In this, the inflaton phenomenology used to implement inflation, is
remarkably successful.  Though our intention is to adhere as closely
as possible to QFT and GR to explain, or rather to interpret
cosmogenesis and the inflaton scenario, inevitably, unconventional
notions will appear.  For the inflatonic phenomenology unfolds at the
planckian scale, and even transplanckian when introducing the primeval
fluctuations born during inflation as modes, in order to explain
present day observations which reflect them.  And physics at these
scales we do not have.  Nevertheless, we shall hazard hypotheses which
are based on accessible physics.  This procedure follows past
traditions in our science, hypotheses elaborated from what is known,
through analogy, to attempt to elucidate the unknown.  For each
fundamental advance carries with it its own limitations. To overcome
these limitations, such hypotheses, first elaborated, are then
incorporated into a new logical structure which transcends the past.
At the present time, that limitation is met at the planckian scale.

It is the nature of our enquiry to reject the idea of an elementary field
having the peculiar properties requisite to having a successful inflaton.
Likewise, we eschew a priori constructs which exist beyond the pale of
conventional physics, string theory, supergravity, extra dimensions. The
particular hypotheses we adopt are, in great measure, forced upon us by
pursuing conventional physics to the limits of its validity.

We shall dwell further on the concept of causality applied to cosmogenesis.

Inflation, hence creation, in GR, is naturally born in a planckian seed from
which both the modes of QFT and the classical geometric structure of GR are
simultaneously created, together with the geometry. No seed, no geometry. No
geometry, no physics as we know it, for physics as we know it requires a
geometric background to support the propagation of the modes of QFT, and modes
are the ``ur-stuff" of physics.

The conclusion is that physics is born as the universe is. At present, our
limited intellectual horizons, built, as they are, upon QFT and GR forces upon
 us a phenomenology of creation and precludes a complete theoretical formulation. The
tools necessary come by a rational account of pre-cosmological chaos,
unendowed, as
 it is, with geometry, bereft of the modes of QFT, are simply absent from the arsenal of
weapons which appear essential to formulate a satisfactory rational account of creation.

Were there no such
homogeneous structuralism of space, brought about by the hubble
expansion, then there would be no modes of QFT, hence no positive
energy to drive the hubble expansion which is brought about by the GR
energy constraint [$H^2=\rho/3$; $H=$ hubble expansion parameter,
$\rho=$ vacuum energy density]. The conclusion is that physics is born
when the universe is, at least physics as we know it.

Nevertheless, the conjectured nature of the planckian seed of creation is in
great measure the substance of this paper. These conjectures will lead to a
formulation of the inflaton, the nature of its substance in terms of fields,
and its mass, which is more intelligible than the more familiar phenomenology
as expressed, for example, in [1]. The conjectural character is based on
analogy with known physics, for example, the role of quantum tunnelling out of
chaos.  As such it is still phenomenology.

The hope is that, in the future, there will emerge a coherent mathematical
scheme, a rational account of our world and the transplanckian pre-world from
which our world has sprung. Call it quantum gravity, a discipline which will be
called upon to describe chaotic configurations without the support of geometry,
but which lays the foundations of geometry on the macroscopic scale (and
possibly even at a cisplanckian scale only slightly greater than planckian).
Loop quantum gravity affords some hope, but unfortunately there has not yet
been forthcoming a large scale geometry that would emerge from the embryonic
notions that form the canonical formulation of LQG.

Thus we must abide by our ignorance and live with a phenomenological,
yet suggestive, account of creation from ``formless void''.  This
paper is to be read as a suggestive interpretation of a highly
successful phenomenology.

To appreciate the unfamiliar and enigmatic issues that arise in inflatoniary
phenomenology we first introduce the subject with a brief synopsis.

The inflaton is a scalar field, $\phi$, whose dynamics is governed by a
covariantized Klein-Gordon (KG) equation, the metric for which is an expanding
flat homogeneous space. The hubble expansion parameter, $H(\phi,\dot{\phi})$,
is self-consistently determined in virtue of the GR energy constraint,
$H^2/3=m^2\phiav^2/2 + \langle\dot{\phi}\rangle^2/2$. We set $m_{pl}=1$, and
also treat other irrelevant constants as equalling $1$. The averaging,
indicated by $\langle\,\rangle$, is over the homogeneous expanding patch. If
$\phi$ is not homogeneous, it rapidly becomes so, in virtue of the expansion,
for a wide class of initial conditions. The phenomenology is most often
presented with initial conditions $\langle\dot{\phi}\rangle = 0$,
$\langle\phi\rangle \ne 0$. Once more, if these conditions are not met
initially, there is an attractor solution  of the KG equation which rapidly
brings about the ``slow roll'' condition $\langle\dot{\phi}\rangle \ll
m\langle\phi\rangle$, hence a quasi stationary hubble parameter whereupon the
homogeneous patch expands quasi-exponentially.

 The KG equation has an in-built friction term, which results from the
expansion itself, so $\langle\phi\rangle \Rightarrow 0$ after a few $e$-folds
of the scale factor, $a$.  For the homogeneous patch to expand sufficiently to
account for the homogeneity of space within our present horizons, determined by
the adiabatic expansion subsequent to inflation, one requires initially
$\phiav=O(10)$, a numerical result that follows from the use of the mass, $m$
of $\phi$ determined observationally to be $O(10^{-6})$.

The condition $(\phiav/ m)\gg 1$, required for the slow roll, is amply
fulfilled and that is the sine qua non for inflation to work. [Often
the term $m^2\phi^2$ is replaced by a more general function $V(\phi)$,
but the quadratic approximation for $V(\phi)$ seems to suffice.]

During the period called ``reheating'', as $\phiav \to 0$, the energy stocked
in $\phiav$, principally in the potential energy $m^2\phiav^2$, gets converted
into quanta through a presumed coupling of $\phi$ to the conventional fields of
QFT.  The mechanism of coupling and the resulting dynamics is less well
understood than the rest.  But one postulates that, in the end, there is
thermalization of the resulting quanta and the adiabatic expansion ensues. Thus
inflation prior to reheating, and reheating itself, is the phenomenology that
replaces what was the big bang prior to the invention of inflation.  It is not
only causal, but it has physical consequences which are very interesting and
observationally confirmed, to wit: the formation of primeval fluctuations
behind the so-called inflationary horizon at $H^{-1}$.  Upon following the
evolution of these fluctuations throughout cosmological history, one shows that
their traces survive to this day.  They account both for the large scale
distribution of galaxies as well as the characteristically scale-dependent
temperature fluctuations of the cosmic microwave background (CMB).  The
analysis is quite general and its success is a triumph of no small
accomplishment.

From this synopsis, the principal conclusion is that inflation is a consequence
of the dynamics of the gravitation/field complex which unfolds at the planckian
level. Moreover, nothing prevents extrapolation to its beginning wherein the
scale factor is of planckian dimensions. Hence inflation is an invitation to
new physics.

The principal physical ingredient which we call upon to explain cosmogenesis
and inflation is intra-field gravitational interactions.  At short length
scales the resulting negative energy density dominates the more familiar zero
point energy density. The latter is often cited as the problem of the
cosmological constant, which is the problem of how one must tune that component
of energy to get finite reasonable physics. Such a consideration, however,
neglects to take into account the intra-field gravitational energy.  Whereas
zero point energy density roughly scales like $l^{-4}$ at length scale $l$
(from $\sum^{l^{-1}}_{\vec{k}}|k|$ in free field theory), the latter in a
perturbative scheme is expected to scale roughly like $l^{-6}$ (in units where
$l_{pl}=m_{pl}^{-1}=1$).  For example a Newtonian estimate is
$\int^{l^{-1}}\!\!d^3k \int^{l^{-1}}\!\!d^3k' |\vec{k}| |\vec{k}'| / |\vec{k}
-\vec{k}'|^2$.

We suppose that on scales $l > l_{pl}$, a more precise perturbative
treatment will not give a result qualitatively different.

Guided by common sense as well as the mode description of field configurations,
presumably a guide of some reliability at scales $l > l_{pl}$, we may suppose
that the number of field degrees of freedom scales like $l^{-3}$. Thus, save
for special circumstances we surmise that field configurations are
characterized by behaviour at small $l$. But, for $l < l_{pl}$, gravitational
interaction energy within a field configuration is dominant. If so, this would
render unnecessary the introduction of a subtraction term in the form of a
cosmological constant to counter infinite zero point energy.

Let us now go further along these lines. The energy constraint of GR, applied
to homogeneous flat space, reads $H^2 = \rho/3$.  From the previous paragraph
this implies that, save special circumstances, cosmology as we view it (i.e.,
as an expanding large scale homogeneous space which is the habitat of field
configurations) cannot be realized.  This is because, as argued, save special
circumstances, configurations are dominated by the small $l$ scale for which
gravitational interaction energy dominates, where $\rho< 0$.

The problem of cosmogenesis is thus the search for special circumstances
wherein field configurations with scale $l > l_{pl}$ become relevant.  We shall
argue that this happens when a mechanism that leads to homogeneity sets in, and
this happens in virtue of a very particular tunnelling mechanism out of the
chaotic small $l$ configurations to those of larger $l$.  Specifically, let us
denote by $\Lambda_0$ that momentum scale at which $\rho(\Lambda_0)=0 $ so
that, for momenta $\Lambda$, for which $\Lambda < \Lambda_0$, we have
$\rho(\Lambda)>0$.

Then, the configuration
of larger length scale towards which one tunnels is at the momentum scale
$\Lambda\approx\Lambda_0-m$, where $m \ll 1$ in planckian units. In
that configuration, there arises approximate local homogeneity, a
planckian seed slightly larger than planckian in volume.  Within that
seed, configurations start to get sorted out to form modes, giving
rise to $\rho > 0$, hence a local hubble expansion.  Inflation sets in
and field configurations similar to those within the seed also
arise, this time being stimulated by the hubble expansion whose effect
continues to extend from the seed to its neighbours. Thus $a$ expands
exponentially and a universe is born.

From whence arrive these small length scale degrees of freedom of
negative energy in which the energy has small absolute value?
Note that in their sequestered state they do not contribute to the hubble
expansion until they tunnel out to be converted to positive energy
configurations whereupon they then do contribute to the local hubble
expansion.  They arise as quantum fluctuations, or if one will, as
virtual configurations of positive energy.  Under normal circumstances
these virtual configurations which tunnel out of sequestration would
simply reenter into the ensemble of the sequestered configurations
giving rise to no net effect.  However, gravity in the form of
the hubble expansion seizes those fluctuations, so as to stabilize them,
converting virtuality to reality. That is the working of the energy
constraint $-H^2 + \rho/3 = 0$.

It remains to model the mechanism of sequestration.  We take
sequestered configurations to be bound state space-time entities which
are localized within the causet sites of Sorkin.  This picture is an
interpretation of [4] wherein dark energy is given as a
vacuum fluctuation.  It is adopted here because it has the
virtue of Lorentz invariance, and moreover has met with some
success in accounting for present day dark energy, albeit it
requires refinement to deal with dark energy at an earlier epoch of
the adiabatic expansion! One may envision other models
of sequestration such as foam on a space-like surface, but
it would seem that Lorentz invariance is a good guide.  In this respect,
see Section 3 where the causet scenario is briefly presented.

From this rather naive model, one can start to estimate the parameters of the
inflaton scenario (Section 5).  The inflaton is taken to be a collective degree
of freedom formed from the positive energy degrees of freedom of all elementary
fields, which lie in the momentum range ($\Lambda_0-m$, $\Lambda_0$), with
$\rho(\Lambda_0)=0$.  The parameter, $m$, is the inflaton's mass.  We will
estimate from the tunnelling mechanism, as roughly $O(10^{-5})$, as against the
phenomenologically favored value of $O(10^{-6})$.  This is a satisfactory
preliminary result which is an encouraging sign.  From this value of $m$, and
further argumentation in Section 5, one can estimate the amplitude $\phiav$ of
the inflaton condensate: $\phiav \approx O(\sqrt{N})$ where $N$ is the number
of species of fields contributing to known (cisplanckian) physics.  Thus
$\phiav=O(10)$, which is once more a phenomenologically acceptable estimate,
wherein the requirement to have sufficient inflation to accommodate the
``size'' of our present universe is instrumental.  Further arguments are given
in Section 5 wherein the causal set discretization of space-time indicates that
$\phi$ is a scalar field governed by a Klein-Gordon evolution operator
appropriately coarse-grained.  The condensate $\phiav$ is the average of $\phi$
over the homogeneous expanding patch.

Also, energetic considerations impose the creation of a charge neutral
universe.  Therefore we distinguish matter anti-matter neutrality and
charge neutrality.  From the above, the scenario which is proposed
would result in a universe containing only matter degrees of freedom,
but which is charge neutral, as is required by observation.

While we note that fermionic fields tend to contribute negatively to the
vacuum energy, we will here postpone a detailed discussion of the implications.

We conclude this introduction with a summary of the issues that will
be presented in more detail than is indicated in the introduction.
\begin{enumerate}
\item{$\phiav=O(10)$ is a sort of condensation of $\phi$, that emerges
    at creation, formed from modes in the momentum interval
    $(\Lambda_0 -m)< |k| <\Lambda_0)$ where $\rho(\Lambda_0)=0$).
    $\phiav^2$ may be viewed as a superfluid density where the fluid
    is composed of matter.}
\item{$\phi$ obeys a KG equation wherein figures the mass, $m
    =O(10^{-5})$, this estimate resulting from a tunnelling mechanism
    from sequestration into a homogeneous positive energy density
    configuration.  The d'Alembertian in the KG operator arises from a
    coarse grained description of propagation on a causet.}
\item{Primeval fluctuations arise behind the inflationary horizon at
    $O(H^{-1})$ through essentially the same mechanism as $\phiav$
    arises, save that it is their wave vector that characterizes them
    rather than mass.  What is important in both cases is the creation
    of cisplanckian modes, not quanta.  This occurs through tunnelling.
    In the text, comparison is made with the similar situation that
    occurs in black hole evaporation.}
\end{enumerate}

\noindent For the facility of the reader the Table of Contents is:
\begin{enumerate}
\item{Introduction} \item{Negative Pressure and Inflation} \item{The Cis-Trans
Dichotomy of Field Theory} \item{Causal Set} \item{Cosmogenesis and Tunnelling}
\item{Brief Comments on Fluctuations and Black Hole Evaporation}
\end{enumerate}

\section{Negative Pressure}

There are interesting points in common between field theory and fluid
dynamics.  These are often fruitful when searching for the dynamics of
fields in the cosmological setting and will be called upon throughout
this paper.

This section is devoted to the strange concept of negative pressure.
It was encountered in the earliest work on inflation [2,3] when it was
noticed that the exponential expansion of space, which was postulated
to be the causal prelude to the big bang was characterized by a
negative pressure when expressing the energy-momentum tensor
$T_{\mu\nu}$ of a field in terms of fluid variables, with $p =$ pressure
and $\rho =$ mass density.  To see this, one begins with $T_{\mu\nu}$
for a field expressed in terms of its 4 velocity, $u_\mu$ and one
replaces $u_\mu$ by the field momentum $\partial_\mu\phi$.
Identifying the tensors gives the interpretation of the field
expression for $T_{\mu\nu}$ in terms of pressure:
\begin{eqnarray}
  \rho = \frac{1}{2}\phi^{,\mu}\phi_{,\mu} + V(\phi)\\
  p = \frac{1}{2}\phi^{,\mu}\phi_{,\mu} - V(\phi)  
\end{eqnarray}

It is readily seen that if $V(\phi) > 0$ and dominates the
derivative terms, then the space expands exponentially; and
$p<0$.

Does this formal result express some physical content? It does and it
is suggestive of how one must proceed to model inflation.

The simple fluid setup to illustrate $dE=-pdV$ is to calculate the work done on
a fluid enclosed within an insulated cylinder one end of which is a piston
which exerts pressure on the fluid.  The pressure is due to forces external to
the fluid, taken positive if they are exerted inwards so that $dV<0$. Then
$p>0$ assures $dE = -pdV > 0$ due to work done on the system.  Mutatis
mutandis, if $dV>0$ the systems works on the piston and $dE<0$, the principle
of the steam engine which does work in its expansion phase.

For the cosmological fluid during inflation, one has $dV>0$ and $p<0$
so that, under the cosmological expansion, $dE>0$. Apparently, there
are sources of energy within the system which push the system out so that
energy within the fluid increases as the volume does.  These sources
must be rather homogeneously distributed so that there is not too much
dilution of these internal energy sources as the system expands. One
wants inflation to last for a sufficiently long time to assure the
existence of a universe at least as large as ours.  Therefore the
total number of sources must increase in number.  And if the expansion
is quasi-stationary (called the slow roll corresponding to a very slow
decrease of the density of potential energy), their density should
remain quasi constant. Thus, the expansion is exponential.

This roughed out macroscopic sketch of inflation indicates the way
towards a more detailed mechanism. One must search for energy sources
within field configurations which, like the field itself, are
homogeneously distributed on the coarse-grained level.

The phenomenological KG equation containing a mass term, and
covariantized, offers a successful rendering of what is required.  In
virtue of the attractor solution, $\phiav$ varies little on a time
scale $O(\phiav_{initial}/m)$ maintaining the energy density constant
during this period at $\rho\approx m^2\phiav^2_{initial}$ whereupon
the system then exits from inflation; i.e., $\phiav\to 0$.  After this
``reheating'' period, the inflationary energy density degenerates into
quanta, hence temperature and entropy.

\section{The Cis-Trans Dichotomy}

The action of a scalar field in flat non-expanding homogeneous space is
$\int d^4x \partial_\mu\chi \partial^\mu\chi$.  Masslessness, in the
context of this paper, is an adequate approximation since we shall be
interested in dynamics at the planck scale and all known masses are
$\ll 1$.  Flat homogeneous space is used for the sake of simplicity.
Its use is vindicated at a later stage by virtue of inflation itself.

Let $\nu(\Lambda)$ denote the number of modes, cut-off at $\Lambda$, and let
$\rho(\Lambda)$ denote the energy density. Then, in the absence of gravity, one
has:

\begin{equation}
  \nu(\Lambda) = \Lambda^3
\end{equation}
\begin{equation}
  \rho(\Lambda) = \Lambda^4
\end{equation}

The absence of an a priori length scale allows one to let $\Lambda\to\infty$
with impunity since the absence of gravity implies zero coupling of
$\rho(\Lambda)$ to anything.

All is changed when gravity enters the game, i.e., when one adds a term
$\int\sqrt{g} R d^4x$ to the action and makes the field
action invariant converting it to $\int
\sqrt{g}\partial_\mu\chi\partial^\mu\chi\,d^4x$ where $g_{\mu\nu}$ is
used to make $\partial_\mu$ and $\partial^\mu$ covariant and
contravariant.  It is beyond present
knowledge to calculate $\rho(\Lambda)$ as $\Lambda \to 1$, but the
Newtonian approximation and more generally its refinement in
perturbation theory (in powers of $m_{pl}^{-2}$) offer a guide.  The
Newtonian approximation gives for the interaction energy density

\begin{eqnarray}
  \rho(\Lambda)|_{\stackrel{Newtonian}{^{Interaction}}} &=&
 -\int^\Lambda\!d^3k \int^\Lambda\!d^3k'  |\vec{k}| |\vec{k}'|
/ |\vec{k} -\vec{k}'|^2 \nonumber\\
  &=& -\Lambda^6 ~,\\
  \rho(\Lambda)|_{_{Newtonian}} &=& \Lambda^4 - \Lambda^6 ~.
\end{eqnarray}

To cisplanckian scales (i.e., $\Lambda<1$) a perturbative estimate
would be expected to yield the form
\begin{equation}
  \rho(\Lambda)|_{Pert.}=\alpha(\Lambda)\Lambda^4-\beta(\Lambda)\Lambda^6
  ~, \label{e37}
\end{equation}
where $\alpha$ and $\beta$ are slowly varying positive functions. There is no
need here to enter into complications of renormalization. But we do specify
that the word ``perturbative'' implies the prescription that all momentum
integrals are cut off at scales $\Lambda = O(1)$.  For one of the principal
theses of this paper is that perturbative QFT, based as it is on the modes of
the fields, is inadequate at planckian and transplanckian scales.

In subsequent sections it shall be argued that straightforward perturbative QFT
makes sense only at scales $\Lambda <\Lambda_p$, where $\Lambda_p=\Lambda_0-m$;
where $\rho(\Lambda_0)=0$.  For brevity, in this section we ignore the
interesting subtleties (which are essential to describe inflation) for
$\Lambda$ near $\Lambda_0$ and simply class modes with $(\vert \vec{k}\vert
<\Lambda_0)$ as cis.  They propagate and the trans degrees of freedom ($\vert
\vec{k}\vert>\Lambda_0$) do not, nor do they contribute to the energy
constraint ($H^2=\rho/3$) since $\rho<0$ for $\Lambda>\Lambda_0$.

We now formulate the postulate that only the cis sector of field configurations
determines the space-time geometric back reaction occasioned by those
configurations. Thus, the hubble expansion parameter is determined from the cis
configurations.  And analogously, the mass, M, of a black hole which varies due
to evaporation is determined (directly!) only by the cis sector.  The
qualification ''directly'' is further explained below.

In similar fashion we expect dark energy, if it is indeed a vacuum
effect as most physicists believe, to be directly influenced by cis
configurations.

Our postulate is most reasonable. Consider that $\rho(\Lambda)$ goes negative
for $\Lambda > \Lambda_0$ in consequence of the gravitational attraction
responsible for the $\Lambda^6$ term of Eq.6. For $\Lambda>\Lambda_0$, the fact
that $\rho(\Lambda)$ goes negative in the perturbative calculation indicates
that the transplanckian degrees of freedom are localized in bound states.  It
is this fact that vitiates transplanckian perturbation theory.

We postulate that the transplanckian class of field configurations are
aggregates of bound states, each aggregate of length scale
$\Lambda_0^{-1}$.  This postulate goes back to the foam idea of
Wheeler of the mid-60's.  It has been given a Lorentz invariant
setting in more recent times in terms of Sorkin's causet scenario [4].
Since such configurations are sequestered and do not extend beyond
length scale $O(\Lambda_0^{-1})$, it is natural to postulate that
their influence cannot extend over macroscopic regions of space-time,
hence the postulate: only cis configurations influence macroscopic GR.

But that does not mean that the transplanckian configurations are inert.  On
the contrary, the main content of Section 4 is based on cis-trans
communication.  Whence, it will be argued, the mass, m, in the KG
equation for the cis-class concept $\phi$, arises in consequence.
This is the origin of the indirect influence of the trans class in cis
physics.

An alternative way to show up the cis-trans dichotomy is to be found
in an important series of papers of Parentani [5] who has shown, by
perturbative methods, that modes become overdamped at the planckian
scale.  Thus they become useless to characterize
transplanckian field configurations.

Parentani's calculations, being perturbative, do not indicate the
destiny of overdamped modes.  The calculation must be followed up non
perturbatively, one suggestion for which are the bound state
aggregates which we have called upon above to give an effective
discretization of space (space-time).

The model of discretization which will serve as a basis for our
discussion of inflaton dynamics is the causet construction.  It has
the advantage of explicit covariance, i.e., propagation by discreet
movements (hopping) on the causet network is, on the average,
covariantly described.  For ``on the average'' we shall sometimes use
the term coarse grained, referring to an average over scales O(1). The
causet scheme is designed for the purpose.

This is not to say that a non-covariant scheme such as foam lain out on
space-like surfaces could not equally well serve the purpose, and in fact, even
be more suggestive of physical models.  This is Parentani's view.  It is for
definitiveness and convenience that we adopt causets.

For the non cognoscenti a brief summary of the causet scenario follows. The
prescription is sketchy and mathematically non rigorous.  For more details, see
[4].

\section{Causets}

Space-time is here given as a manifold upon which we distribute points.  For
simplicity we deal with 1 spatial dimension. Generalization to 3 dimensions is
straightforward, see [6].  Also, we shall use flat space-time.  This suffices
to describe the space-time patches we call upon for the description of
propagation in the cosmological spaces we envision.

Each point in the manifold is the site of an event; hence it can serve as the
tip of a light cone.  Signals are sent from within the cone to the tip.  The
causet prescription is to distribute these points at random in a Poisson
distribution.  Then under Lorentz transformation not only are the boundaries of
the cone invariant, but so are the lengths ($\Delta t^2 - \Delta x^2 = \Delta
u\Delta v$) between events where $u, v=t\mp x$.  That is because Poisson
distributions are lain down with equal areas (on average) subtended by the
parallelograms drawn with $\Delta u, \Delta v$ on a side.  (The Minkowski
version of the distribution of raindrops on a flat pavement is a convenient
visualization.)  Since under Lorentz transformations, space-time intervals
transform as $\Delta v \to \Delta v\,\exp[w]$ and $\Delta u \to \Delta u
\exp[-w]$ where $w$ is the rapidity, a coarse grained view of the distribution
is conserved.

From these, Sorkin and collaborators have constructed a mean
d'Alembertian.  It is the inverse of a retarded Green's function, this
latter constructed as a hopping matrix from site to site in the
forward direction.  The ingenious rules for its construction make it
sufficient to go backwards from the tip of the light cone to sites
a few units of length in the past.  There results, on taking the
average of the inverse, a Lorentz invariant d'Alembertian.

This ``tour de force'' lays the foundation of a field theory possessed
of a Lorentz-invariant cutoff.  In [6], details on the propagator
are further developed and it is hoped that shortly we will have a
full-fledged QFT based on this construction.  In Section 5 we make use
of this knowledge to rationalize the KG equation that governs the
dynamics of $\phi$.

\section{Cosmogenesis and the Inflaton}

We now assemble the considerations of the previous sections. To start
we discuss the formation of the inflaton condensate $\phiav$, then
how one can rationalize the value of its mass.

The basic assumption is that the inflaton field emerges from
sequestered degrees of freedom, which, for definiteness we take to be
bound state aggregates within causet elements. $\phi$ is a collective
field comprised of the momentum components of all fields contributing
to conventional (cisplanckian) QFT. $\phiav_t$ is the spatial average
of $\phi$ at time t over the expanding homogeneous patch of space that
is described by inflation.  The momentum components have $\vec{k}$
in the interval ($\Lambda_0-m < |\vec{k}| < \Lambda_0$) where
$\rho(\Lambda_0)=0$ defines $\Lambda_0$ and $m\ll\Lambda_0$ .  $m$
will be identified with the inflaton mass.

First we devote ourselves to the question of how to rationalize
$\phiav_{t_0} = O(10)$, where $t_0$ is the initial time of inception
of inflation, here taken to be the time of cosmogenesis.

Let $\nu(\Lambda)$ be the number of all modes up to scale $\Lambda$. For
$t<t_0$, $\nu(\Lambda)=0$.  There are no modes of QFT, all field configurations
being sequestered and they all have negative energy density, $\rho(\Lambda)<
0$, so there is no expansion.  In fact, as emphasized in Section 1, there is no
physics, at last in the sense as we conventionally formulate physics.

Now consider a single causet element.  Field degrees of freedom tunnel in and
out as fluctuations.  Tunnelling is important when their configurations have
nearly vanishing energy density.  We may think of virtual transitions wherein
negative energy configurations spend a finite duration of time exterior to the
main bulk of the causet element having virtual positive energy.  Here we appeal
to analogy with familiar quantum physics.  That is all we have to go on.

Under normal circumstances these virtual configurations will fall back
into their sequestered state leaving no net effect.  The exceptional
circumstance resulting in a planckian seed of inflation arises when
that fluctuation has a sizable homogeneous component of energy over a
volume slightly greater than planckian.  This will result in a local
hubble expansion whose effect will extend in an isotropic manner to
the neighboring causet elements.  In this way the cooperative process
gets set up as originally envisioned in [1] wherein curvature
drives positive energy field fluctuations to give permanent effects,
and these positive energy effects drive curvature.

To make quantitative progress, let $\nu(\Lambda(x,t))$ be the number of modes
which issue from the causet site at $x,t$ in a manner that is approximately
homogeneous and isotropic and which is localized around $x,t$.  Then the
momentum components contributing to $\nu(\Lambda(x,t_1))$ will be planckian in
character (since causet elements are planckian distributed). Since
$\nu(\Lambda(x, t_1))$ is composed of cisplanckian field configurations, the
corresponding energy carried by them is positive as well as
[$\Lambda_0-\Lambda(x,t)] > 0$.  [We recall $\rho(\Lambda_0)=0$ and
$\Lambda(x,t) < \Lambda_0$ to make $\rho(\Lambda(x,t)) > 0$.]  It then makes
sense to define the positive energy quantity $\phi^2$ through
\begin{equation}
  m\phi^2=\nu(\Lambda_0) - \nu(\Lambda(x,t)) ~, \label{e51}
\end{equation}
and the spatial (homogeneous) average of $\phi$
\begin{equation}
  \phiav^2_t = [\nu(\Lambda_0) -
  \langle\nu(\Lambda(x,t)\rangle]\,/\,m ~. \label{e52}
\end{equation}

\noindent where $\langle\nu(\Lambda(x,t))\rangle$ is the spatial mean of the
number of modes contained on the cloud of cisplanckian field configurations
centered about $t$.  The quantity $m$ is a mass parameter and
$\langle\Lambda(x,t)\rangle = \Lambda_0 - m$ with $m\ll\Lambda_0$.

In Eq. \ref{e51}, one may interpret $\nu(\Lambda(x,t))$ as a fluid
density, that fluid being composed of the modes of QFT.
$\nu(\Lambda)$ is the number of modes of all species having $|\vec{k}|
< \Lambda~~(=\Lambda^3$ for free field theory).  Then to good
approximation
\begin{equation}
  \phiav^2_0 \approx \frac{d\nu}{d\Lambda}|_{_{\Lambda_0}} \approx
\frac{\nu(\Lambda_0)}{\Lambda_0}. \label{e54}
\end{equation}
$\phiav_t$ is to be interpreted as the inflation condensate wave
function.  It is the analog of a superfluid wave function at absolute
zero since the fluid carries no entropy.  The factorization of
superfluid density into a square is the analog of the Penrose-Onsager
construction.

That $\phiav^2_t$ is indeed the homogeneous part of the phenomenological
inflaton is seen by multiplying Eq.\ref{e52} and integrating over $d^3x$ in the
volume about the site $x,t$.  The conventional potential energy $\int
m^2\phiav^2_td^3x$ that is stocked in that volume is then checked out to be
equal to $m\int[\nu(\Lambda_0) - \nu(\Lambda(x,t))]d^3x$.  This latter is the
number of modes that have leaked out into the volume multiplied by the energy,
$m$, carried by each of those modes.

The value of $\phiav_t$ can be estimated from Eq. \ref{e54}.  Taking
$\Lambda_0=O(1)$ one has
\begin{equation}
  \phiav_t=O(\sqrt{N}) ~. \label{e55}
\end{equation}

At $t_0$, one requires phenomenologically $\phiav_{t_0} = O(10)$. Since
$N=O(10^2)$ the result conforms to the phenomenological requisite.

The next step is to estimate $m$.  The \emph{sine qua non} for inflation to
take place is the homogeneous isotropic production of energy per tunnelling
event.  Were energy produced helter skelter from chaos, without giving rise to
a homogeneous distribution of cisplanckian field energy, we would not be able
to avail ourselves of the GR energy constraint and induce a global hubble
expansion.  Rather such a random event would result in an uneventful
fluctuation, i.e., $\phiav = 0$, $H=0$.

We now interpret $m$ in terms of the product of the energy involved in each
tunnelling event and the amplitude, $A$, for a tunnelling to occur. For the
former we take the energy $O(1)$, since that is the only scale present.  For
the latter, in order for a tunnelling event to set up the cascade of successive
events which take place from neighbor to neighbor, the tunnelled wave function
must have extension out to the distance of neighbour also $O(1)$, in virtue of
the causet construction.  Thus without any further additions, if such a
tunnelling event were responsible for inflationary cooperative process, we
would have $m=1e^{-1}$ where the exponent comes from $e^{-\Delta E\,d}$ with
$\Delta E=$ planck energy and $d=$ mean interval between causet elements.  But,
as mentioned above, to set up the tunnelling events necessary for inflation,
the tunnelled configuration should approximate to an isotropic homogeneous
distribution.  Therefore tunnelling out of a given causet site should be a
synchronized act of tunnelling from a given site to neighbors.  If there are
$Z$ in number then one obtains $m=O(e^{-Z})$.  In three dimensions $Z=O(10)$
whence $m=O(10^{-5})$ as against the phenomenological value $O(10^{-6})$. Thus
this rough estimate once more conforms to phenomenology in order of magnitude.

It is well to emphasize that we are still working within a
phenomenological framework making use of analogies borrowed from
conventional physics.  The hope is that these encouraging estimates
render accepted phenomenology somewhat more intelligible, but even
more that they will direct further research towards a complete
theoretical structure.

Further development of inflatonionary physics requires an understanding of how
the inflaton propagates.  Fortunately, propagation of fields on the causet
network has been the subject of recent work which complements previous work of
Sorkin and collaborators [6].  We shall not review this work in the present
paper, but do hope in the future to return to a detailed consideration of the
hopping mechanism in terms of the localized field configurations which have
been posited to exist at the causet sites.  One should not overlook the
possibilities of interesting dynamics involving cis--trans communication as
part of hopping, whence possible interesting asymmetric effects.  For the
nonce, combining the conceptual understanding of $\phi$, $m$ and propagation,
there exists some element which makes inflaton physics intelligible.

In this paper we shall not touch upon reheating.  Once $\phiav_t =
O(1)$, after a few $e$-folds, the KG equation predicts that
$\dot\phi$ and $m\phi$ become comparable. Thereupon the slow roll and
inflation stop; the universe is made and there is no more significant
creation.  Interaction of the cis degrees of freedom among themselves,
gravitationally and otherwise, cause $\phi$, a coherent collective
degree of freedom, to fall apart.  Analogy with fluid dynamics points
to the transition between laminar and turbulent flow as a possible
theoretical point of departure to seek inspiration.  Another
possibility is a quantum formalism of the disintegration into myriads
of lower energy quanta.  For all we know this may be another way to
look upon turbulence.

The problem of the generation of primeval fluctuations presents an
interesting construct to the creation of inflationary energy.  Yet the
fundamentals are the same, the latter being the creation of
homogeneous configurations out of sequestrations.

Suppose the seed of cosmogenesis is at length scale $a$, where
$a=O(1)$, which then increases to macroscopic size.  A fluctuation
of comoving wave vector $\vec{k}$ comes into existence when the
physical wave vector, $\vec{k}/a$ is $O(\Lambda_0)$ whereupon it
emerges from sequestration into the cis world.  As a expands, so do
the values of $\vec{k}$, so as to give a scale invariant system of
observable fluctuations.  The birth of fluctuations is thus the
analogy of the birth of $\phi$, mode by mode, without
condensation for their energy does not participate significantly to
the homogeneous Hubble expansion.  An account of the density of these
fluctuations is the content of [1].

\section{Concluding Remarks}

The essential message of this paper is the existence of a length
scale, $\Lambda_0^{-1}$, below which the energy density at scale
$\Lambda^{-1} ~~(=\rho(\Lambda))$ becomes negative.  The most
important consequence is that, save for exceptional circumstances, the
field configurations do not expand, owing to the constraint
$H^2=\rho/3$.  There are exceptional circumstances which permit field propagation.
Namely, this can occur when the field is $O(\sqrt{N})$.

The exceptional circumstances which result in a cosmos through inflation, arise
when an ordered cisplanckian field configuration of positive energy density
$\rho(\Lambda)$, with $\Lambda < \Lambda_0 $, tunnels out of sequestration. The
requirement of homogeneity whence the subsequent existence of the modes of QFT,
is met by the existence of a mean scalar field, $\phiav$, the inflaton and
$\rho(\Lambda) = m^2\phiav^2$, where $\phiav$ is the mean of $\phi$ over the
expanding homogeneous patch and $m$ is the mass of the inflaton.

$\phiav$ arises from a seed, emerging from chaos by tunnelling out of a bound
state configuration that is centered on a causet site.  If one makes the
natural hypothesis that the energy scale of binding is $O(1)$ and the
inter-causet length scale is $O(1)$ as well, then the requirement that this
seed yield an approximately homogeneous field distribution concomitantly with
an approximately homogeneous space surrounding that seed, yields a value for
$m$ consistent with the requirements of observation: $m=O(10^{-6})$.  This
follows from the existence of roughly 12 neighbors about each site.  The same
interpretation of the $\phi$ field in terms of mode density and mass yields the
estimate $\phiav = O(\sqrt{N}) = O(10)$ where $N$ is the number of species of
fields contributing to QFT.

The second main subject of this conclusion concerns the parallelism
between black hole evaporation, dark energy and cosmic creation.

Degrees of freedom, which are destined to become Hawking evaporated quanta,
near to the black hole horizon at $r=2M$, exhibit a strong blue shift and enter
into the transplanckian regime.  Therefore, unless otherwise solicited, they
will remain sequestered on the causet sites near the horizon.  However, as for
cosmogenesis they give rise to fluctuations which tunnel out of sequestration.
Being cisplanckian at this point they can elicit macroscopic response, in this
case the change of mass of the black hole metric.  As for the cosmic problem,
one effect drives the other, ultimately leading to permanent effects wherein
the frequency of asymptotic Schwarzschild space-time is equal to the loss of
black-hole mass as expressed in the metric change.  It is interesting that the
dynamics of both black hole evaporation and cosmic creation of quanta requires
such elaborate mechanisms: the Bogoljubov transformation for Kruskal to
Schwarzschild modes on the former, inflation and the formation of the inflaton,
followed by reheating in the latter.

A final point concerns dark energy, where hopefully the considerations
above can be substantiated in the presence of cosmic
inhomogeneous fluctuations as well as more complicated distributions
of on mass shell quanta than were dealt with in the literature.  If
successful, there is hope that all processes concerning creation from
``formless'' transplanckian void can be united.  Whereupon we shall
have some guidance for constructing the fundamental theory.  The concept of
unitarity will have to be modified, at least in the sense of dealing
with constantly varying cisplanckian Hilbert spaces.
\bigskip\newline
\bf Acknowledgements: \rm  Very helpful discussions are acknowledged with Achim
Kempf, Rafael Sorkin and Martin Green.
$$$$
\subsubsection*{Bibliography}

\begin{itemize}

\item[{[1]}] V. Mukhanov, \it Physical Foundations of Cosmology, \rm Cambridge University Press, Cambridge (2005)

\item[{[2]}]
R. Brout, F. Englert, E. Gunzig, \it The creation of the Universe as a quantum phenomenon, \rm, Ann. Phys. (USA), {\bf 115}, 78 - 106 (1978)

\item[{[3]}]
A. A. Starobinsky, \it A new type of isotropic cosmological models without singularity, \rm Phys. Lett. {\bf B91}, 99-102 (1980)

\item[{[4]}]
L. Bombelli, R.K. Koul, J. Lee, R. Sorkin, \it Quantum source of entropy for black holes, \rm Phys. Rev. {\bf D34}, 373 (1986).

\item[{[5]}] R. Parentani, \it Confronting the trans-Planckian question of inflationary cosmology with dissipative effects, \rm Class. Quantum Grav. {\bf 25} 154015 (2008)

\item[{[6]}] S. Johnston, \it Feynman Propagator for a Free Scalar Field on a Causal Set, \rm
 Phys. Rev. Lett. {\bf 103}, 180401 (2009)

\end{itemize}
  \end{document}